# PARAMETER ESTIMATION FOR COMPUTATIONALLY INTENSIVE NONLINEAR REGRESSION WITH AN APPLICATION TO CLIMATE MODELING

By Dorin Drignei,[1] Chris E. Forest[2] and Doug Nychka[1]

*Oakland University, Massachusetts Institute of Technology and Pennsylvania State University, and National Center for Atmospheric Research*

Nonlinear regression is a useful statistical tool, relating observed data and a nonlinear function of unknown parameters. When the parameter-dependent nonlinear function is computationally intensive, a straightforward regression analysis by maximum likelihood is not feasible. The method presented in this paper proposes to construct a faster running surrogate for such a computationally intensive nonlinear function, and to use it in a related nonlinear statistical model that accounts for the uncertainty associated with this surrogate. A pivotal quantity in the Earth's climate system is the climate sensitivity: the change in global temperature due to doubling of atmospheric $CO_2$ concentrations. This, along with other climate parameters, are estimated by applying the statistical method developed in this paper, where the computationally intensive nonlinear function is the MIT 2D climate model.

**1. Introduction.** A fundamental question in understanding the Earth's climate system is quantifying the warming of the atmosphere due to increased greenhouse gases. This relationship is formalized by the climate sensitivity, a parameter defined as the increase in global mean surface temperature due to a doubling of $CO_2$ in the atmosphere. Although climate sensitivity and other climate parameters are informed by observations, their impact can only be evaluated by simulations of climate with a numerical computer model. Such a model usually includes atmosphere, ocean, land and ice components and is called an atmosphere ocean general circulation model

Received December 2007; revised September 2008.
[1]Supported by NSF Grant DMS-03-55474.
[2]Supported in part by the NSF-CMG Grant DMS-04-26845.

(AOGCM). Because climate is defined as a long term average of weather, an AOGCM is usually run (integrated) over many years in order to establish its mean behavior. Thus, numerical experiments with these models require extensive computational resources and the number of runs (also termed integrations) is often limited. For example, the Community Climate System Model (CCSM) requires months of time on a supercomputer to simulate a few hundred model years. Typically an AOGCM depends on unknown parameters which need to be estimated and the statistical challenge is to estimate these parameters along with companion measures of uncertainty using a limited set of climate model experiments.

The general statistical problem addressed here is parameter estimation in a nonlinear regression model [e.g., Seber and Wild (1989)], where the nonlinear regression function is computationally intensive to evaluate, such as an AOGCM. In particular, the example discussed here involves observed climate data and the MIT 2D climate model [Sokolov and Stone (1988)] as a nonlinear function of three uncertain parameters collectively denoted by $\theta$: the equilibrium climate sensitivity $S$, the diffusion of heat anomalies into the deep ocean $K_v$ and the net aerosol forcing $F_{\mathrm{aer}}$. Here we take a maximum likelihood approach and pay particular attention to the correlation structure and its effects on the uncertainty measures of the resulting estimates. The standard nonlinear regression approach for this particular estimation problem assumes the following statistical model for the observed climate data:

$$(1) \qquad \mathbf{Y} = \mathbf{f}_\theta + \varepsilon,$$

where the errors are assumed normal with zero mean vector and covariance matrix $W$. The estimated parameters of this statistical model, including $\theta$, are then obtained by maximizing the likelihood

$$(2) \qquad \left(\frac{1}{\sqrt{2\pi}}\right)^N \frac{1}{\sqrt{\det W}} \exp\left\{-\frac{1}{2}(\mathbf{Y}-\mathbf{f}_\theta)' W^{-1}(\mathbf{Y}-\mathbf{f}_\theta)\right\}.$$

This is usually achieved by using an iterative algorithm. Notice, however, that this requires computing $\mathbf{f}_\theta$ for many values of $\theta$, or, equivalently, running the climate model for a possibly large number of $\theta$ values. Such an approach is not feasible for applications where $\mathbf{f}_\theta$ is an AOGCM or even the simplified MIT 2D climate model. To overcome this computational difficulty, we will substitute a statistical surrogate for $\mathbf{f}_\theta$, denoted $\widetilde{\mathbf{f}}_\theta$, which will result in a much faster estimation algorithm for the unknown parameters $\theta$. The statistical model for the observed data is now

$$(3) \qquad \mathbf{Y} = \widetilde{\mathbf{f}}_\theta + \mathbf{E}_\theta + \varepsilon,$$

where $\mathbf{E}_\theta$ is the error in approximating $\mathbf{f}_\theta$ by $\widetilde{\mathbf{f}}_\theta$ and $\varepsilon$ is observation error. More precisely, we use analysis of computer experiments methodology [e.g.,

Santner et al. (2003)] to analyze climate model output data at a sample of 'input' parameter vectors and build a statistical surrogate to predict the climate model output at new, untried parameter values. This analysis is empirical Bayesian in its nature [as described in Currin et al. (1991) for scalar output and in Drignei (2006) for multiple outputs], in the sense that a Gaussian process serves as a prior distribution for the multiple outputs and the posterior distribution is used to predict the output at new parameters. We take $\tilde{\mathbf{f}}_\theta$ to be the posterior mean, but also use the uncertainty about this mean to adjust the likelihood of observations. To our knowledge, the empirical Bayesian analysis of multiple output computer experiments, used in the context of physical model calibration through maximum likelihood methods, is new. The statistical model we develop also accounts for different types of uncertainty (e.g., climate model internal variability), it includes correlation with respect to various dimensions (e.g., space–time correlation) and a new feature of our work is to include the uncertainty of the surrogate model as part of the estimated parameter uncertainty. The calibration of computationally intensive computer models has been recently done mostly through fully Bayesian models, for example, Kennedy and O'Hagan (2001), Craig et al. (2001), Bayarri et al. (2007), Higdon et al. (2008) and the last two references dealing with multidimensional computer model outputs.

We make use of the data sets from Forest et al. (2002) who implemented a Bayesian model for the same problem and have shown results from flat and expert priors (see top row of Figure 2). A notable feature from a climatological point of view in their work is the skewness of the climate sensitivity $S$, which appears to be more pronounced for flat priors. Appendix A presents a geophysical argument in support of this skewness. Forest et al. (2003), Forest, Stone and Sokolov (2006) and Sanso et al. (2008) revisited and refined the Bayesian approach. Several other studies estimated a probability density function for climate sensitivity [Andronova and Schlesinger (2001), Gregory et al. (2002), Knutti et al. (2002, 2003)]. All such studies are based on estimating the degree to which a climate model can reproduce the historical climate record. More recently, this same technique has also been applied to the climate record for the past 600 years [Hegerl et al. (2006)]. Last Glacial Maximum (LGM) climate change can also be used as an additional line of evidence to estimate the probability density function of the climate sensitivity and such results have been combined by Annan and Hargreaves (2006) to provide a more complete picture of available constraints for placing bounds on climate sensitivity.

This paper is organized as follows. Section 2 discusses the climate observations, the climate model and the output data sets. Section 3 develops the statistical surrogate for the climate model, while Section 4 shows how this surrogate can be used to construct a computationally efficient statistical model for the climate observations. The results are presented in Section 5 and the paper ends with some conclusions in Section 6.

**2. Observed and modeled climate.**

2.1. *Climate observations.* In this analysis we use three separate sets of observations representing the ocean heat content, surface temperatures and upper air temperatures.

The oceans play an important role in the planet's climate system because they can store and transport large amounts of heat. The subsurface ocean temperature records are sparse and contain the most uncertainty due primarily to the difficulty in obtaining such temperature data sets. The data used in this paper originate in the Levitus et al. (2000) ocean temperature data set from the surface through 3000-meter depth, showing a net warming over the last half of the twentieth century.

The upper-air temperature data set used in this paper originates in the improved upper-air temperature data recorded by radiosondes and described in Parker et al. (1997). The later data are considered somewhat more reliable than satellite-based Microwave Sounding Units (MSU) data because they provide a longer record and a better vertical resolution. Nevertheless, the MSU data have proved to be useful in removing time-varying biases of radiosonde temperature data caused, for example, by changes of instrumentation or operating procedures. Parker et al. (1997) describe some interpolation, bias- and error-removing methods for the original sparse and irregular radiosonde upper-air temperature data sets.

Among all the temperature records, the surface temperatures are the longest, most spatially complete and documented. This is due primarily to the existence of meteorological stations throughout the world for a relatively long time. The data sets used in this paper originate in the extended and interpolated data sets of Jones (1994).

Summary statistics for each observational dataset are used to make the appropriate comparison with the climate model (as discussed later). The averaging methods for each diagnostic are discussed in Forest et al. (2002) and serve to remove the short time-scale variability that is not associated with the long time-scales changes in climate. These averages result in the patterns summarizing the changes in mean temperatures for each source as discussed in Section 2.3.

2.2. *The MIT 2D climate model and the unknown parameters.* AOGCMs are the primary tools for predicting changes in global climate patterns on planetary scales and at decadal or longer time scales. Mathematically, these models are systems of partial differential equations derived from the laws of fluid dynamics and thermodynamics for the atmosphere, ocean, ice and land systems, in three spatial dimensions (3D) and a temporal dimension. Since these models are nonlinear, they are usually solved numerically over a space–time grid. These numerical methods, however, require large computational resources and, therefore, have limited flexibility for exploring parametric (or structural) uncertainty. Most often, for a given AOGCM, individual parameters are set according to performance of individual components (e.g., cloud parameterizations or ocean sub-grid scale mixing parameterizations) and then held fixed or modified in a heuristic fashion when the fully coupled AOGCM is assembled and tested. For a more statistical approach to determine parameters, we require a flexible climate model designed to explore uncertainty in the large-scale response (e.g., global or hemispheric average temperature changes) in the fully coupled system. The MIT 2D climate model [Sokolov and Stone (1988)] is one such model designed for these purposes and was used in this research. Some technical aspects of the MIT 2D climate model are given in Appendix B.

The MIT 2D model has two parameters that determine the decadal to century response to external factors that drive the climate. These are the equilibrium climate sensitivity $S$ (measured in degrees $K$) to a doubling of $CO_2$ concentrations and the global-mean vertical thermal diffusivity $K_v$ (measured in $cm^2/s$) for the mixing of thermal anomalies into the deep ocean. Sokolov and Stone (1988) have shown that the large-scale response of AOGCMs can be duplicated by the MIT 2D model with an appropriate choice of these two parameters. This correspondence supports the study of the climate system using the simpler MIT model because it can approximate more complex and realistic three dimensional models. The third parameter considered in this paper is the net aerosol forcing strength $F_{\mathrm{aer}}$ (measured in $W/m^2$) and controls the amount of cooling of the atmosphere to increased amounts of particles. The computational time required to simulate 50 years with the MIT 2D climate model is about 4 hours on a 3 GHz Pentium 4 Linux workstation and it is several orders of magnitude faster than simulation with a state-of-the-art 3D AOGCM. Given its flexibility to duplicate AOGCM responses and its computational efficiency, the MIT 2D climate model is a tool well suited for answering questions which would be impractical to explore with 3D AOGCMs.

2.3. *Specific data sets and model output.* Due to the sparsity and the large uncertainties in the deep-ocean temperature data, only the linear trend in the observed temperature record is retained for analysis so that the ocean observed data is just a scalar. The upper-air temperature observations are the difference in the 1986–1995 and 1961–1980 mean temperatures, recorded at each 5° latitude and at 8 pressure levels (850 hPa through 50 hPa). In order to simplify the analysis, 10 latitude coordinates containing mostly missing data have been discarded, so that the final upper-air temperature change observed data set is a $26 \times 8$ matrix. The surface temperature data set has been averaged so that the final surface temperature change data set is a $4 \times 5$ matrix, corresponding to 4 latitude bands by 5 decades.

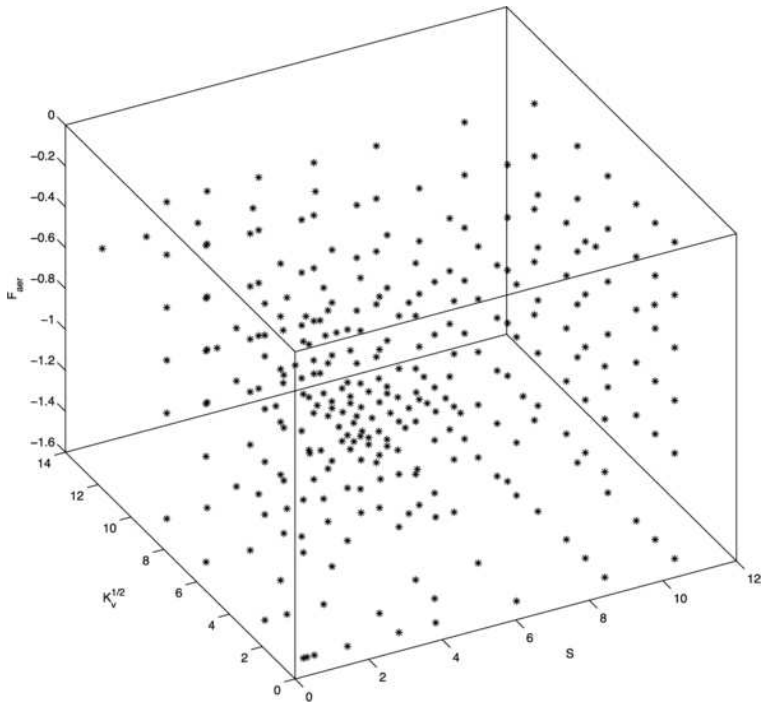

FIG. 1. *The sampled inputs.*

Let $\theta = [S, K_v^{1/2}, F_{\text{aer}}]$ be the parameter vector. We considered $D = 306$ parameters $\theta_i$ as in Forest et al. (2002), sampled in a parameter space (see Figure 1) to cover a range of parameters well beyond the domain corresponding to current AOGCMs. In Forest et al. (2002), an initial, approximately factorial design over the domain has been selected and then a second, higher density sampling has been added in the region of highest likelihood to better estimate their entire distribution. For each parameter vector in this parameter space, the relatively large model output data set is transformed so that its format matches that of the observed data sets: the deep-ocean temperature trend (scalar), the upper-air temperature changes (26 latitudes and 8 levels) and the surface temperature changes (4 latitude zones and the last 5 decades of the last century). There are four replicates for each of these data sets, called ensemble members, obtained by changing the initial conditions in the climate model. In this paper we work with averages across ensemble members, although the ensemble variability will be accounted for in our statistical model.

**3. Statistical surrogate for the climate model.** The statistical model for observations developed in this paper has two components. The first component is a surrogate of the climate model output over the region of model parameters and the second component connects the surrogate with the observed data. Each of these components will be discussed next.

3.1. *Statistical model for climate model output.* All three observational data sets have a similar statistical model, so we will only present the model for surface temperature changes in detail. The climate model surface temperature change can be conveniently organized as an array over time, space and climate parameters, with dimension $N_T \times N_Z \times D$. (For our data sets, $N_T = 5$, $N_Z = 4$, $D = 306$.) To better describe the statistical model, this three-dimensional data set is stacked as a vector and simply denoted by $\underline{\mathbf{f}}$. Climatology arguments suggest that the numerical climate model be decomposed into two random components: $\underline{\mathbf{f}} = \underline{\mathbf{f}}^s + \underline{\mathbf{f}}^n$ corresponding to a climate signal $\underline{\mathbf{f}}^s$ and noise process $\underline{\mathbf{f}}^n$ ascribed to climate model internal variability.

The correlation matrix for the signal $\underline{\mathbf{f}}^s$ can be written, under the computationally convenient assumption of separability, as a Kronecker product of smaller correlation matrices $C_\Theta \otimes C_Z \otimes C_T$ reflecting the various dimensions of the problem. The power exponential [Sacks et al. (1989)] was chosen as a simple but flexible model to describe the correlations among each dimension of climate signal. The matrix $C_\Theta$ is defined as

(4) $$C_\Theta(\theta, \theta') = \exp\left(-\sum_{i=1}^{3} \eta_i |\theta_i - \theta'_i|^{p_i}\right).$$

The correlations among zonal bands and across time are also assumed to follow the power exponential family, although analysis based on the Matérn family [e.g., Stein (1999), page 31] has also been carried out. The correlation matrix for the noise $\underline{\mathbf{f}}^n$ will be $I \otimes \Gamma$, with a consistent empirical estimator of the space–time covariance $\Gamma$ obtained by assuming the same space–time structure across the $D$ sampled parameters and 4 ensemble members (after subtracting the ensemble mean at each of the $D$ sampled parameters). A simple method based on wavelet basis functions [Nychka et al. (2002)] was used to estimate a nonstationary form that takes advantage of the assumption that $\Gamma$ does not depend on $\theta$ and can be thought of as a regularization of the sample covariance matrix. The space time fields are expanded in a W-transform multiresolution basis, with $\mathcal{W}$ being the matrix of wavelet basis functions that relate coefficients to the model field and $\mathcal{D}$ being the sample covariance matrix for the coefficients. $\mathcal{D}$ is decomposed as $\mathcal{D} = \mathcal{H}^2$ and the elements of $\mathcal{H}$ are thresholded by setting 90% of the elements to zero, resulting in a regularized matrix $\tilde{\mathcal{H}}$. We use the estimate $\Gamma = \mathcal{W}\tilde{\mathcal{H}}^2\mathcal{W}'$. Then

$$\Sigma_\Theta = \sigma^2(C_\Theta \otimes C_Z \otimes C_T) + \omega^2(I \otimes \Gamma)$$

is the overall covariance matrix, so that $\underline{\mathbf{f}} \sim \mathbf{N}(\mu\mathbf{1}, \Sigma_\Theta)$. The output data analyzed here are temperature changes and averages, therefore, we assume that any large scale trends have been eliminated, justifying the choice of a statistical model with constant mean. An unbiased estimate of the parameter $\omega^2$ is obtained by pooling the ensemble sample variances across inputs, space and time. The remaining statistical parameters are estimated by the maximum likelihood method and all parameters are fixed at their estimated values throughout the rest of the statistical analysis.

3.2. *The statistical surrogate and its error.* The statistical model described above is used to construct a surrogate for the climate model at an arbitrary parameter vector $\theta$ in the parameter space. To define the surrogate for $\mathbf{f}_\theta$ and its error, we consider the conditional distribution of the climate signal $\mathbf{f}_\theta^s$ on the climate model output data $\underline{\mathbf{f}}$, which is multivariate normal of mean vector

$$\tilde{\mathbf{f}}_\theta = \mu \mathbf{1} + \tilde{\Sigma}_{\theta\Theta} \Sigma_\Theta^{-1} (\underline{\mathbf{f}} - \mu \mathbf{1})$$

and covariance matrix

$$V_\theta = \sigma^2 (C_Z \otimes C_T) + \omega^2 \Gamma - \tilde{\Sigma}_{\theta\Theta} \Sigma_\Theta^{-1} \tilde{\Sigma}'_{\theta\Theta},$$

where $\tilde{\Sigma}_{\theta\Theta} = \sigma^2 (C_{\theta\Theta} \otimes C_Z \otimes C_T)$ and $C_{\theta\Theta}$ is defined as in (4). The surrogate is $\tilde{\mathbf{f}}_\theta$ and its error $\mathbf{E}$ has a multivariate normal distribution $\mathbf{N}(\mathbf{0}, V_\theta)$. The surrogate model for the upper air temperature is defined similarly. The surrogate model for the deep ocean temperature trend is much simpler because it is based on a univariate response.

## 4. Likelihood function for the observed data.

4.1. *Likelihood function for a single data set.* Here we develop the model for the observed surface temperatures, whereas Section 4.2 presents the complete statistical model for all three data sets.

The statistical surrogate developed above will not predict perfectly the climate model output at new parameters, nor do we expect it to match perfectly the observed data. Therefore, if $\mathbf{Y}$ denotes generically the observed data set in vector format, then the corresponding statistical model is

$$\mathbf{Y} = \tilde{\mathbf{f}}_\theta + \mathbf{E}_\theta + \varepsilon,$$

where $\varepsilon$ is observation error with multivariate normal distribution $\mathbf{N}(\mathbf{0}, \tau^2 R_Z \otimes R_T)$ and the elements of the matrices $R_Z$ and $R_T$ are power exponential correlations with parameter vector $\boldsymbol{\xi}$. Here $\mathbf{E}$ and $\varepsilon$ are assumed independent. An important statistical point is that the discrepancy between $\mathbf{f}_\theta$ and $\tilde{\mathbf{f}}_\theta$ is modeled explicitly through the analysis of the computer experiments' approach in the previous section. The likelihood of the model for $\mathbf{Y}$ is

$$
\begin{aligned}
L(\mathbf{Y}|\theta,\tau,\xi) = &\left(\frac{1}{\sqrt{2\pi}}\right)^{N_Y} \frac{1}{\sqrt{\det(V_\theta + \tau^2 R_Z \otimes R_T)}} \\
&\times \exp\left\{-\frac{1}{2}(\mathbf{Y} - \tilde{\mathbf{f}}_\theta)'(V_\theta + \tau^2 R_Z \otimes R_T)^{-1}(\mathbf{Y} - \tilde{\mathbf{f}}_\theta)\right\}.
\end{aligned}
$$
(5)

Whereas the computationally intractable likelihood (2) contains the nonlinear climate model $\mathbf{f}_\theta$ and the covariance matrix $W$ of the observation error, the more computationally efficient likelihood (5) depends on the surrogate $\tilde{\mathbf{f}}_\theta$ of the climate model and a covariance matrix based on two components: surrogate error and observation error.

4.2. *Full likelihood.* Let $\mathbf{Y}_s, \mathbf{Y}_o, \mathbf{Y}_u$ denote the observed surface temperature change, the observed deep ocean temperature trend and the observed upper air temperature change respectively. These three data sets are assumed independent conditioned on the true climate signal at $\theta$ so that the overall likelihood to be optimized is a product of the likelihood functions developed in the previous subsection, for the three specific data sets:

$$L(\mathbf{Y}_s, \mathbf{Y}_o, \mathbf{Y}_u|\theta,\tau_s,\tau_o,\tau_u,\xi_s,\xi_u) = L(\mathbf{Y}_s|\theta,\tau_s,\xi_s)L(\mathbf{Y}_o|\theta,\tau_o)L(\mathbf{Y}_u|\theta,\tau_u,\xi_u).$$

(Notice that the observation error for the ocean temperature trend is univariate normal.) This likelihood takes about 5 seconds to be evaluated in Matlab on a computer with dual 2.6 GHz Xeon processors and 4 GB RAM.

There are geophysics arguments in support of a right skewed distribution for the climate sensitivity parameter (see Appendix A) and the literature on the estimation of this parameter reports various degrees of skewness. This, in turn, makes it difficult to agree on an upper confidence bound for this important parameter: how large will the global mean surface temperature be when the $CO_2$ amount is doubled? Here we would like to investigate this aspect through the finite sample distribution of the maximum likelihood estimators. An established method for such purpose is the parametric bootstrap [Efron and Tibshirani (1993)], where synthetic data $\bar{\mathbf{Y}}_s, \bar{\mathbf{Y}}_o, \bar{\mathbf{Y}}_u$ will be generated from the multivariate normal of likelihood $L(\mathbf{Y}_s, \mathbf{Y}_o, \mathbf{Y}_u|\hat{\theta},\hat{\tau}_s,\hat{\tau}_o,\hat{\tau}_u,\hat{\xi}_s,\hat{\xi}_u)$, with the maximum likelihood estimates taken as "true" values of the parameters. The new likelihood of the simulated data will be maximized and the point estimate $(\bar{\theta},\bar{\tau},\bar{\xi})$ will be obtained. This process is repeated $B$ times and, therefore, $B$ independent simulated values $(\bar{\theta},\bar{\tau},\bar{\xi})$ will be obtained, which will further be used to summarize the distribution of $(\hat{\theta},\hat{\tau},\hat{\xi})$, for example, through confidence regions. A direct search method in Matlab has been used to optimize the likelihood functions throughout this paper.

**5. Results.** The statistical model described in Section 3.1 has been fitted to the output data and the parameter estimates for the power exponential correlations given in Table 1. An important component of the statistical model is the choice of covariance family for the statistical surrogate. To determine the sensitivity to the power exponential correlation, the Matérn family was also considered and we present how the inference on the climate parameters is effected by this alternative model for correlations. Table 1 indicates smoothness in the latitude–time dimensions for surface temperatures and latitude–pressure dimensions for upper-air temperatures. Therefore, Matérn correlations generating one-time mean square differentiable realizations (i.e., with the smoothness parameter $\nu$ set to 1.5) have also been fitted in those dimensions, with the correlation parameter estimates shown in Table 2. More specifically, the statistical model for the output data based on the Matérn correlations is identical to the model presented in Section 3, except that the elements of the matrices $C_Z$ and $C_T$ are now Matérn instead of power exponential correlations, with temporal and latitude range parameters denoted by $\alpha_t$ and $\alpha_s$, respectively. The power exponential maximized loglikelihood minus the Matérn maximized loglikelihood is 97 for the surface temperature and 1555 for the upper air output data sets, respectively. The models with these correlations have been further used to build the surrogates and the associated errors as described in Section 3.2. These have been used in the likelihood of the observed data (Section 4). $B = 300$ bootstrap simulations were obtained, which have been further used to obtain nonparametric kernel marginal density estimates and confidence intervals/regions. The results are summarized in Figure 2. The upper row contains the univariate marginal density estimates with 99% confidence intervals for both power exponential (solid) and Matérn (dot) models, as well as posterior pdfs from Forest et al. (2002); it appears that a mild right skewness of the estimated densities for the climate sensitivity $S$ is present. The lower row contains the kernel density estimates of the bivariate marginal densities under the power exponential (solid) and Matérn (dot) models. The contours correspond to 90%, 95% and 99% levels confidence regions. The results from these two correlation models (power exponential and Matérn) do not appear to be largely different. The inclusion of correlations was helpful in better constraining the climate system parameters than in the statistical model considered in Forest et al. (2002), which did not include such correlations.

TABLE 1
*Estimates of parameters in the statistical model for output data (the power exponential model): surface temperature (upper row), upper air (middle row) and deep ocean (lower row)*

| $\mu$ | $\eta_1$ | $p_1$ | $\eta_2$ | $p_2$ | $\eta_3$ | $p_3$ | $\eta_t$ | $p_t$ | $\eta_s$ | $p_s$ | $\sigma^2$ | $\omega^2$ |
|---|---|---|---|---|---|---|---|---|---|---|---|---|
| 0.217 | 1.365 | 0.425 | 1.189 | 0.273 | 2.283 | 0.903 | 1.227 | 1.147 | 1.255 | 1.501 | 0.024 | 0.016 |
| −0.047 | 1.206 | 0.302 | 1.031 | 0.185 | 1.274 | 0.390 | 14.476 | 1.849 | 4.382 | 1.431 | 0.007 | 0.004 |
| 0.001 | 3.430 | 1.999 | 21.636 | 1.999 | 1.545 | 1.927 | — | — | — | — | $2 \times 10^{-6}$ | $7 \times 10^{-9}$ |

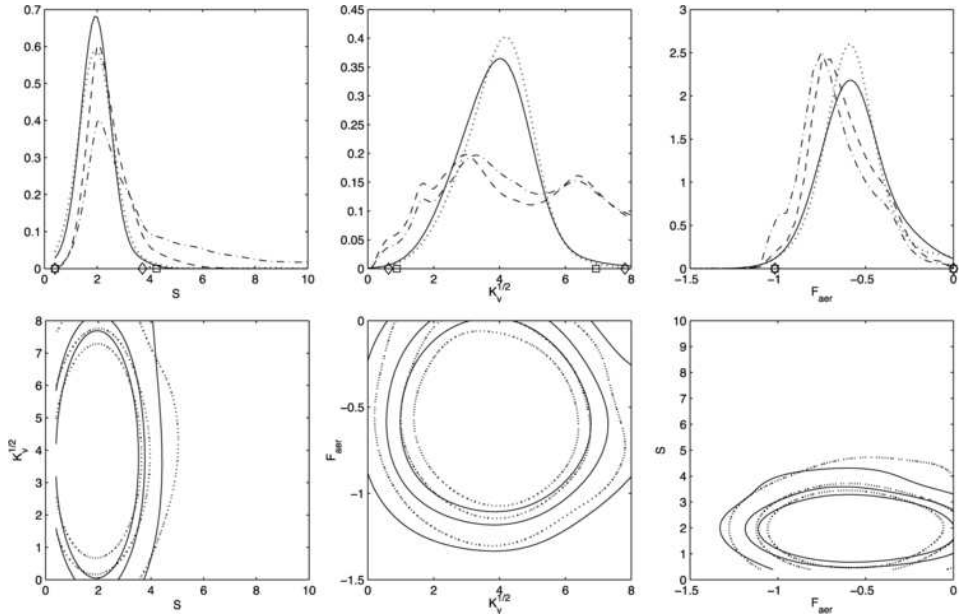

FIG. 2. *Upper row: estimated univariate marginal densities based on power exponential correlations (solid) with 99% confidence intervals (diamond), and on the Matérn correlations (dot) with 99% confidence intervals (square), and the posterior pdfs from Forest et al. (2002): expert prior for climate sensitivity (dash) and flat priors (dash–dot); Lower row: estimated bivariate densities with contours representing 99%, 95% and 90% confidence regions (solid lines: power exponential; dotted lines: Matérn).*

TABLE 2
*Estimates of parameters in the statistical model for output data (the Matérn model): surface temperature (upper row), upper air (lower row). The deep ocean estimates and $\omega^2$ are the same as in Table 1*

| $\mu$ | $\eta_1$ | $p_1$ | $\eta_2$ | $p_2$ | $\eta_3$ | $p_3$ | $\alpha_t$ | $\alpha_s$ | $\sigma^2$ |
|---|---|---|---|---|---|---|---|---|---|
| 0.211 | 1.399 | 0.426 | 1.239 | 0.267 | 2.287 | 0.908 | 3.987 | 2.682 | 0.022 |
| $-0.021$ | 1.142 | 0.323 | 0.963 | 0.193 | 1.171 | 0.376 | 6.873 | 12.244 | 0.006 |

**6. Conclusion.** This paper has presented a computationally efficient statistical model as an alternative to a naive nonlinear regression model involving a computationally intensive nonlinear function. Our approach is to construct a statistical model that includes a faster running surrogate for the computationally intensive nonlinear function and it is underlined by some general principles. The statistical model offers a comprehensive framework to accommodate various sources of uncertainty. For example, we incorporated an error term to represent climate internal variability and a component for uncertainty in the surrogate, as well as an observation error term. The covariance matrices are full rank and estimated from the available forced model output data. The statistical model proposed accounts for correlation in all dimensions of the problem (e.g., it includes spatio-temporal correlation). This strategy was helpful to better constrain the unknown climate model parameters through tighter confidence intervals and regions, especially for the climate sensitivity $S$ and ocean diffusivity $K_v$ parameters.

## APPENDIX A: CLIMATE SENSITIVITY

An important characteristic in the Earth's climate system is the climate sensitivity $S$: the change in global temperature due to doubling of atmospheric $CO_2$ concentrations. In complex climate models derived from the fundamental equations of physics, the individual processes interact and the feedbacks between the key processes lead to the overall amplification of the temperature change. At the basic level, the additional greenhouse gases (e.g., carbon dioxide, methane and nitrous oxide) have a direct impact on the outgoing longwave radiation in the atmosphere and in the absence of feedbacks (i.e., the nontemperature atmospheric state variables remains unchanged), this change to the system will eventually lead to a given amount of temperature change. But, when the temperature changes, this will inevitably lead to additional changes in the temperature and humidity profiles, cloud distributions, snow and ice cover, and other state variables. These additional changes are the climate system feedbacks that can amplify or reduce the direct impact of the greenhouse gases on the radiative forcing. If we let $\Delta T_o$ be the temperature change due to the direct impact on the radiative transfer, we note that $\Delta T_o$ includes one feedback, the direct impact of temperature change on the infrared spectrum by changing greenhouse gas concentrations, but does not include additional feedbacks from other components of the climate system. It is customary to let the total change in global mean surface temperature above the equilibrium value be written as $\Delta T_{2x} = \frac{\Delta T_o}{(1-\phi)}$ [e.g., see Hansen et al. (1984) or Schlesinger and Mitchell (1987)]. Thus, we have a one-to-one correspondence between the feedback and the climate sensitivity ($\Delta T_{2x}$ and the uncertainties in each can be related).

This relation between climate sensitivity and feedbacks has been known since the early days of climate research and was described in the Charney Report [NRC (1979)], formalized in Hansen et al. (1984) and included in textbooks as early as Henderson-Sellers and McGuffie (1987). Hence, we now have a simple expression to relate the uncertainty in these three variables and from first principles, $\Delta T_o$ and $\phi$ are the two variables which can be quantified individually using climate models and combined to yield $\Delta T_{2x}$. Schlesinger and Mitchell (1987) quantified the uncertainties in feedbacks and recognized that the uncertainties in individual processes will combine to provide an estimate of the total uncertainty. Given the number of processes in the system, it is typical to consider the net feedback to have a normal distribution. However, we also need to consider $\Delta T_o$ as an additional uncertain variable with its own normal distribution, although this is a small contribution compared to $\phi$. By combining these two distributions for $\Delta T_o$ and $\phi$, the resulting distribution for climate sensitivity is expected to be right skewed, although one cannot clearly infer only from these arguments how long the right tail will be.

## APPENDIX B: TECHNICAL ASPECTS OF THE MIT 2D CLIMATE MODEL

The MIT 2D climate model consists of a zonally averaged atmospheric model coupled to a mixed-layer Q-flux ocean model, with heat anomalies diffused below the mixed-layer. The model details can be found in Sokolov and Stone (1988). The atmospheric model is a zonally averaged version of the Goddard Institute for Space Studies (GISS) Model II general circulation model [Hansen et al. (1983)] with parameterizations of the eddy transports of momentum, heat and moisture by baroclinic eddies [Stone and Yao (1987, 1990)]. The model version we use has 24 latitude bands and 11 vertical layers with 4 layers above the tropopause.

The model also employs a Q-flux ocean mixed layer model with diffusion of heat anomalies into the deep-ocean below the climatological mixed layer. This model of the ocean component of the climate system is fully described by Hansen et al. (1983) and only increased computations by a few percent. The model uses the GISS radiative transfer code which contains all radiatively important trace gases, as well as aerosols and their effect on radiative transfer. The surface area of each latitude band is divided into a percentage of land, ocean, land-ice, and sea-ice with the surface fluxes computed separately for each surface type. This allows for appropriate treatment of radiative forcings dependent on underlying surface type such as anthropogenic aerosols. The atmospheric component of the model, therefore, provides most important nonlinear interactions between components of the atmospheric system.

**Acknowledgment.** This research was developed while Dorin Drignei was a visiting scientist at the National Center for Atmospheric Research.

D. Drignei
Department of Mathematics and Statistics
Oakland University
Rochester, Michigan 48309
USA
E-mail: drignei@oakland.edu

C. E. Forest
Department of Earth, Atmospheric
   and Planetary Sciences
Massachusetts Institute of Technology
Cambridge, Massachusetts 02139
and
Department of Meteorology
Pennsylvania State University
University Park, Pennsylvania 16802
USA
E-mail: ceforest@mit.edu
        ceforest@meteo.psu.edu

D. Nychka
National Center for Atmospheric Research
Boulder, Colorado 80307
USA
E-mail: nychka@ucar.edu